\documentclass[%
 aps,
 amsmath,amssymb,
 reprint,%
]{revtex4-1}

\usepackage{graphicx}
\usepackage{dcolumn}
\usepackage{bm}

\usepackage[utf8]{inputenc}
\usepackage[T1]{fontenc}
\usepackage{mathptmx}
\usepackage{xcolor}

\begin{document}

\renewcommand\thesection{\arabic{section}}
\renewcommand{\thesubsection}{\thesection.\arabic{subsection}}
\renewcommand{\thetable}{\arabic{table}}

\title[Sample title]{Absolute standard hydrogen electrode potential and redox potentials of atoms and molecules: machine learning aided first principles calculations}

\author{Ryosuke Jinnouchi}
\affiliation{ 
Toyota Central R\&D Labs., Inc., 41-1 Yokomichi, Nagakute, Aichi, 480-1192, Japan 
}%
\email{jryosuke@mosk.tytlabs.co.jp}
\author{Ferenc Karsai}%
\affiliation{ 
VASP Software GmbH, Berggasse 21, A-1090 Vienna, Austria 
}%
\author{Georg Kresse}
 \altaffiliation[Also at ]{VASP Software GmbH.}
\affiliation{ 
Faculty of Physics and Center for Computational Materials Science, University of Vienna, Kolingasse 14-16, A-1090 Vienna,
Austria
}%

\date{\today}

\begin{abstract}
Constructing a self-consistent first-principles framework that accurately predicts the properties of  electron transfer reactions through finite-temperature molecular dynamics simulations is a dream of theoretical electrochemists and physical chemists. Yet, predicting even the absolute standard hydrogen electrode potential, the most fundamental reference for electrode potentials, proves to be extremely challenging. Here, we show that a hybrid functional incorporating 25\% exact exchange enables quantitative predictions when statistically accurate phase-space sampling is achieved via thermodynamic integrations and thermodynamic perturbation theory calculations, utilizing machine-learned force fields and $\Delta$-machine learning models. The application to seven redox couples, including molecules and transition metal ions, demonstrates that the hybrid functional can predict redox potentials across a wide range of potentials with an average error of 80 mV.
\end{abstract}

\maketitle

\section{\label{section1} Introduction}

The absolute standard hydrogen electrode potential (ASHEP) is the foundation for the thermodynamic measurement of redox potentials. It is defined as the chemical potential of electrons, referenced to the vacuum level, that equilibrates the redox reaction of hydrogen, $1/2\mathrm{H_{2}} \leftrightarrow \mathrm{H^{+}} + \mathrm{e^{-}}$, in its standard state (0.1 MPa for $\mathrm{H_{2}}$ and 1 mol L$^{-1}$ for $\mathrm{H^{+}}$). In many electrochemical experiments, the half-cell potential is scaled to a selected reference electrode, making ASHEP not always a necessary property. However, the absolute potential is a fundamental property that becomes essential when comparing redox potentials to the band edges of metal and semiconductor electrodes or the chemical potential of electrons calculated in electronic structure calculations. ASHEP is also closely related to the absolute value of the real potential of a single ion, which is defined as the free energy change associated with the transfer of an ion from the gas phase to the liquid phase under the standard state. Based on the Born-Haber cycle, the free energy change ($\Delta_\mathrm{at} G^{0}$) of the hydrogen dissociation reaction, $\mathrm{H_{2}} \leftrightarrow 2\mathrm{H}$, and the ionization potential ($\Delta_\mathrm{ion} G^{0}$) of an H atom, allow ASHEP to be converted to the real potential of a proton ($\alpha_\mathrm{H^{+}}^{0}$), which is also referred to as the work function of a proton (here, we use the notations from Ref.~\cite{Trasatti_PAC_1986}).

Despite its importance, determining the ASHEP is highly challenging. The most reliable experimental value is considered to be $-$4.44$\pm0.02$ V recommended by Trasatti and by the International Union of Pure and Applied Chemistry (IUPAC).~\cite{Trasatti_PAC_1986} The value was determined from the potential difference of a voltaic battery.~\cite{Randles_TFS_1956,Farrell_JEAC_1982} However, experimental studies using other methods, such as the Kelvin work function measurement,~\cite{Hansen_JEAC_1979} UHV measurements,~\cite{Thiel_SSP_1987} and cluster-ion solvation data,~\cite{Tissandier_JPCA_1998} have reported scattered results of $-$4.80 to $-$4.28 V.


\begin{table*}[t]
\small
\caption{Real potential of proton ($\alpha_\mathrm{H^{+}}^{0}$) (eV), ASHEP (V) and relevant free energies (eV) calculated by three exchange-correlation functionals (RPBE+D3, PBE0 and PBE0+D3) compared with the experimental values recommended by the International Union of Pure and Applied Chemistry (IUPAC).~\cite{Trasatti_PAC_1986} $\Delta_\mathrm{at} E$ and $\Delta_\mathrm{at} G^{0}$ represent the atomization energy and dissociation free energy of the H${_2}$ molecule, respectively. $\Delta_\mathrm{ion} G^{0}$ is the ionization potential of an H atom in vacuum. MLFF denotes the machine-learned force field trained on the RPBE+D3 data. The specified modeling error bars correspond to $2\sigma$, estimated by block averaging analysis.~\cite{Allen_Book_2008}}
\label{table1}
\centering 
\begin{tabular}{p{23mm} p{25mm} p{25mm} p{25mm} p{25mm} p{25mm}}
\hline
                                                      &MLFF                                   &RPBE+D3                         &PBE0                              &PBE0+D3                            &Exp.    \\
\hline
$\Delta_\mathrm{at} E$                     &4.58                                     &4.58                                 &4.53                                  &4.53                                  &4.73   \\ 
\hline
$\Delta_\mathrm{at} G^{0}$                &4.04                                    &4.04                                  &3.99                                  &3.99                                   &4.21     \\ 
\hline
$\Delta_\mathrm{ion} G^{0}$                &13.75                                 &13.75                                &13.64                                 &13.64                                &13.62 \\ 
\hline
\textbf{$\alpha_\mathrm{H^{+}}^{0}$}     &\textbf{$-$10.98$\pm$0.05}  &\textbf{$-$11.02$\pm$0.06}  &\textbf{$-$11.06$\pm$0.09}  &\textbf{$-$11.12$\pm$0.09} &\textbf{$-$11.28$\pm$0.02} \\ 
\hline
ASHEP                                              &\textbf{$-$4.78$\pm$0.05}  &\textbf{$-$4.75$\pm$0.05}   &\textbf{$-$4.58$\pm$0.09}    &\textbf{$-$4.52$\pm$0.09}  &\textbf{$-$4.44$\pm$0.02} \\ 
\hline
\end{tabular}
\end{table*}

A first-principles (FP) prediction of the ASHEP is also highly challenging. The redox potential $U_\mathrm{redox}$ is determined by the free energy difference $\Delta A$ between the reduced and oxidized states:                                                                                
\begin{align}
U_\mathrm{redox} &= -\frac{\Delta A}{en}, \label{eq_1}                                                                                          
\end{align}
where $e$ is the elementary charge, and $n$ is the number of electrons involved in the reaction. Here, we replace the Gibbs free energy with the Helmholtz free energy, supposing that the changes in volume during the electron-transfer half reaction are negligible. The free energy difference $\Delta A$ can be precisely determined by thermodynamic integration (TI)~\cite{Zwanzig_JCP_1954, Kirkwood_JCP_1935}  In cases where there are significant structural changes from the oxidized form to the reduced form, such as in the hydrogen redox reaction involving solvation and diffusion of the proton, many timesteps are required to accurately determine the free energy difference. Hence, applying the FP calculation directly entails a huge computational cost to achieve good statistical accuracy. Difficulties also arise when evaluating the real potential of a proton in solution. ASHEP is measured with respect to the vacuum level, a quantity that is not directly accessible in simulations using periodic boundary conditions. Furthermore, as demonstrated in previous studies,~\cite{Adriaanse_JCPL_2012, Liu_JPCB_2015, Jinnouchi_npjComputMater_2024} the accurate calculation of redox potentials often requires computationally intensive non-local hybrid functionals that would necessitate several hundred million core hours when complete plane-wave basis sets are used. Therefore, most past calculations have been carried out using approximate methods containing empirical parameters, such as the quantum mechanical molecular mechanical (QM/MM) method,~\cite{Hofer_JCP_2018} static FP calculations assuming water hexamers,~\cite{Tripkovic_PRB_2011} and continuum solvation models.~\cite{Zhan_JPCA_2001, Kelly_JPCB_2006,Jinnouchi_PCCP_2012} Because of the different approximations used the reported simulation values vary from $-$4.56 to $-$4.18 V.

To mitigate the problems, Sprik and co-workers~\cite{Cheng_JCP_2009, Costanzo_JCP_2011, Cheng_PCCP_2012, Liu_JPCB_2015}
introduced a restraining potential that fixes protons to specific water molecules during short, approximately 5-20 picosecond, FPMD simulations to achieve stable and convergent results through TI calculations. They also used localized Gaussian basis basis sets~\cite{Khne_JCP_2020} and norm-conserving pseudopotentials~\cite{Goedecker_PRB_1996}.  Adriaanse and co-workers~\cite{Ambrosio_JCP_2015, Ambrosio_JPCL_2018} used a similar approach with the rVV10 van der Waals functional~\cite{Vydrov_JCP_2010, Miceli_JCP_2015} to calculate the ASHEP at $-4.56$~V, a value that closely matches the  Trasatti's experimental results. However, the use of incomplete localized basis sets can introduce basis set superposition errors. Additionally, the restraining potential used to suppress diffusion in solution might also introduce errors in the entropy of protons. Finally the long first principles calculations remain even today computationally very demanding, limit the statistical accuracy, require state of the art parallel computing facilities, and make routine calculations challenging. Recently, we~\cite{Jinnouchi_npjComputMater_2024} reported that by leveraging machine learning (ML) surrogate models to achieve highly accurate statistical averaging, we can predict the redox potentials of electron transfer reactions of three transition metal redox couples, Fe$^{3+}$/Fe$^{2+}$, Cu$^{2+}$/Cu$^{+}$, and Ag$^{2+}$/Ag$^{+}$, accurately using the PBE0 functional with a fraction of 25 \% exact exchange, projector augmented wave (PAW) method~\cite{Blochl_PRB_1994, Kresse_PRB_1996,Kresse_CMS_1996, Kresse_PRB_1999} and plane waves. However, there are no reports of calculating ASHEP using the all-electron PAW method.

Here, we report the ASHEP and the real potential of protons $\alpha_\mathrm{H^{+}}^{0}$ calculated using FP methods. To achieve sufficient statistics for accurate predictions, we extend the machine learning (ML)-aided thermodynamic integration (TI) developed in our previous study,~\cite{Jinnouchi_npjComputMater_2024} which enabled electron insertion into aqueous solutions, to also allow for proton insertion into aqueous solutions. Using a hybrid functional that includes 25\% exact exchange and dispersion corrections, this method predicts the ASHEP and the real potential of the proton as $-$4.52$\pm$0.09 V and $-$11.12$\pm$0.09 eV, respectively. These values are very close to the IUPAC recommended values of $-$4.44$\pm$0.02 V and $-$11.28$\pm$0.02 eV, as shown in Table~\ref{table1}. In addition to the ASHEP and the three redox couples Fe$^{3+}$/Fe$^{2+}$, Cu$^{2+}$/Cu$^{+}$, and Ag$^{2+}$/Ag$^{+}$ from our previous study,~\cite{Jinnouchi_npjComputMater_2024} we extend applications to the electron transfer reactions of three redox couples, V$^{3+}$/V$^{2+}$, Ru$^{3+}$/Ru$^{2+}$, and O${_2}$/O${_2}^{-}$. The redox couple Ru$^{3+}$/Ru$^{2+}$ forms a rigid first solvation shell composed of six water molecules, similar to the Fe$^{3+}$/Fe$^{2+}$ couple. Past density functional theory calculations using a continuum solvation model~\cite{Jaque_JPCC_2007} have shown that the calculated redox potentials vary by more than 1 V depending on whether the second solvation shell is explicitly included or not, demonstrating the challenges in non-empirically determining solvation structures through static calculations. The redox couple V$^{3+}$/V$^{2+}$ is involved in a half-cell reaction of the redox flow battery.\cite{WeberJApplElectrochm_2011} The non-catalytic electron transfer reaction, O$_2$+e$^{-}$ $\rightarrow$ O$_2^-$, serves as a foundation for elucidating the formation mechanism of the superoxide ion, which is considered the initial precursor in the oxygen reduction reaction in alkaline conditions.~\cite{Anastasijevic_JEAC_1987, Harting_JECA_2002, Ge_ACSCatal_2015} Thus, its FP modeling is considered to be of practical importance. The calculations across these seven redox potentials over a wide range of potentials are expected to demonstrate that our ML-assisted FP method provides a universal framework for accurately predicting redox potentials.

\section{\label{section2}Overview of method}

\subsection{ASHEP and real potential of proton}
We provide a brief overview of our computational methodology for ASHEP. Details are given in Section S1\dag. To facilitate the calculation of the free energy, we divide the hydrogen oxidation reaction into three steps, as in previous studies~\cite{Costanzo_JCP_2011, Le_PRL_2017}: dissociation, H$_{2}$(g)$\rightarrow$2H(g), ionization, 2H(g)$\rightarrow$2H$^{+}+2\mathrm{e}^{-}$(g), and solvation, 2H$^{+}$(g)$\rightarrow$2H$^{+}$(aq), where (g) and (aq) denote species in vacuum and in the aqueous phase, respectively. The corresponding free energy changes are  $\Delta_\mathrm{at} G^{0}$ in the dissociation, $2\Delta_\mathrm{ion} G^{0}$ in the ionization, and $2\alpha_\mathrm{H^{+}}^{0}$ in the solvation. The free energy of the entire redox reaction per electron (defined as $-\Delta A$) is written as follows:
\begin{align}
-\Delta A = \frac{\Delta_\mathrm{at} G^{0}}{2} + \Delta_\mathrm{ion} G^{0} + \alpha_\mathrm{H^{+}}^{0}. \label{eq_2}
\end{align} 
Among the three free energies, $\Delta_\mathrm{ion} G^{0}$ can be easily calculated by a single-point FP calculation. The dissociation free energy $\Delta_\mathrm{at} G^{0}$ can also be easily computed using the ideal gas model. The remaining quantity, the real potential of the proton, $\alpha_\mathrm{H^{+}}^{0}$, can be computed by a TI simulation from the non-interacting proton in the gas phase to the interacting proton in the aqueous phase:
\begin{align}
\alpha_\mathrm{H^{+}}^{0} &= \int_{0}^{1} \left< \frac{\partial H}{\partial \lambda} \right>_{\lambda} d\lambda. \label{eq_3}
\end{align}
Here, $\left<.\right>_{\lambda}$ means evaluation of the expectation value using an ensemble created by the Hamiltonian at coupling $\lambda$. This integral seamlessly connects the proton in the vacuum ($\lambda$=0) to the one in the aqueous phase ($\lambda$=1) along a coupling path. Similarly to the case of the electron insertion method developed in the previous study,~\cite{Jinnouchi_npjComputMater_2024} the Hamiltonian can be written as:
\begin{align}
H &= \sum\limits_{i=1}^{N_{\mathrm{a}}}  \frac{ \left| \mathbf{p}_{i} \right|^{2}}{2 m_{i}} + U_{\lambda}, \label{eq_4} \\
U_{\lambda} &= \lambda \left( U_{1} + e \Delta \phi \right) + \left( 1 - \lambda \right) U_{0}, \label{eq_5} 
\end{align}
where $U_{1}$ and $U_{0}$ denote the potential energy for the aqueous system containing the proton under a periodic boundary condition (PBC) and the one for the pure water and the proton in vacuum under a PBC, respectively.  
In any calculation using periodic boundary conditions, the energy changes due to removal (or addition) of charged species are not entirely well defined. To correct
for this, the potential gap $\Delta \phi$ is introduced. The potential gap essentially specifies the potential of the vacuum level just outside a water surface,
which is the common reference point in electro-chemistry for any charged species (be it electrons or protons). 
There are two alternative views, both of which give the same correction. The potential gap accounts for the chemical potential of the electrons that we move from the gas phase to a reference point just above the surface slab (electron addition to reservoir), or it accounts for the movement of the proton {\it from} 
a reservoir just above the water surface into liquid water. ~\cite{Cheng_JCP_2009, Costanzo_JCP_2011, Cheng_PCCP_2012,Jinnouchi_npjComputMater_2024} Essentially this term fixes the electrostatic reference point of charged species to a point just above the surface of the liquid water.  Consequently, Eq.~(\ref{eq_3}) can be rewritten as:
\begin{align}
\alpha_\mathrm{H^{+}}^{0} &= \int_{0}^{1} \left< U_{1} - U_{0} \right>_{\lambda} d\lambda + e \int_{0}^{1} \left< \Delta \phi \right>_{\lambda} d\lambda. \label{eq_6}
\end{align}
In practice, computing $\Delta \phi$ and the TI in Eq.(\ref{eq_3}) is highly challenging, as relaxation times in water are slow and require expensive ns-scale FP MD simulations that consume millions to tens of millions of core hours. Here, we address these issues by extending the ML-aided scheme developed in our previous study~\cite{Jinnouchi_npjComputMater_2024} from electron insertion to proton insertion.

\begin{figure}
\centering
\includegraphics[width=0.47\textwidth,angle=0]{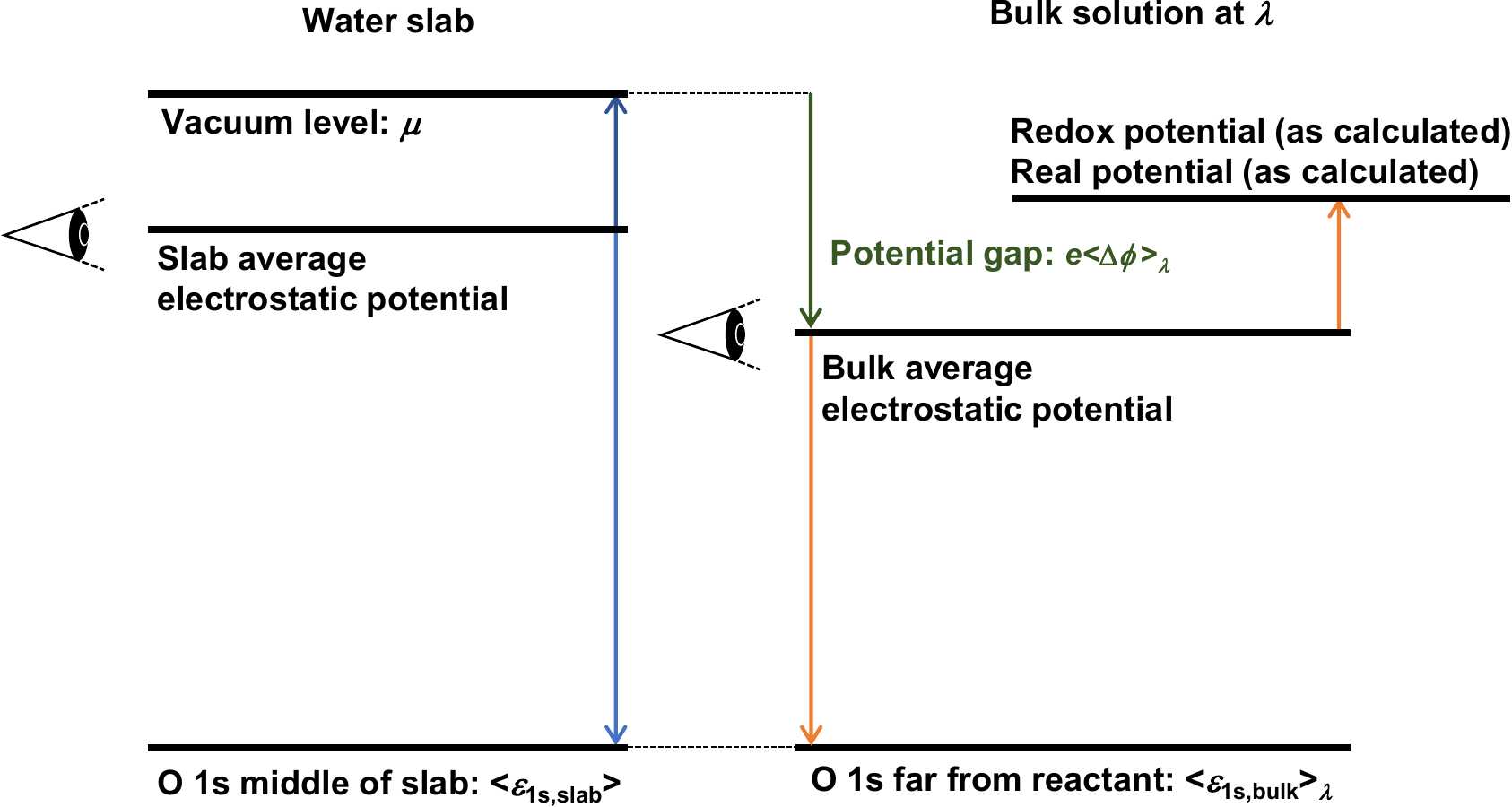}
\caption{Aligning energy levels to the real potential of proton $\alpha_\mathrm{H^{+}}^{0}$. During thermodynamic integration, when a particle with charge $e$ is removed from the slab, the value $\mu$ must be added to account for the fact that the particle is moved to the vacuum level with the electrochemical potential $\mu$. However, this value needs to be aligned with the bulk water calculation. In principle. any reference point can be chosen for the slab and bulk calculations. In periodic boundary codes, it is common practice to set the average electrostatic potential to zero. All eigenvalues and energy changes upon altering the charge state are implicitly referenced to this zero potential point. To correct for the difference in reference points between the slab and bulk calculations, we use the O 1s levels as a common reference point for both systems. The final alignment correction is then obtained as $e \left< \Delta \phi \right>_{\lambda} = \mu - \left< \epsilon_\mathrm{1s,slab} \right> + \left< \epsilon_\mathrm{1s,bulk} \right>_{\lambda}$. 
}
\label{fig1}
\end{figure}

\subsection{Local potential gap}

The local potential gap $\Delta \phi$ needs to be determined to correctly fix the reference potential for charges 
to a point just outside a water surface.  
This reference level is usually determined in a separate slab calculation involving 
an interface between water and the vacuum. But then, how does one align the so determined vacuum level with the periodic slab calculation? 
There are several conceivable options for this procedure. Considering the procedures generally used for semi-conductors,~\cite{Lany_PRB_2008, Freysoldt_RMP_2014} one might be inclined to use the valence band edge (highest occupied orbitals) of water as reference. This choice is not ideal, as the valence band edges are global quantities and might be difficult to identify unambiguously. The averaged local potential ~\cite{Cheng_JCP_2009, Costanzo_JCP_2011, Cheng_PCCP_2012, Le_PRL_2017} has also been suggested as reference level. However, a natural point of alignment are the O 1s levels far from the reactant in the considered periodic cells at any coupling and at the middle of  water slabs, respectively.~\cite{Jinnouchi_npjComputMater_2024}
Similar procedures --- reference points far away from defects --- were also suggested to account for most of the finite size effects~\cite{Lany_PRB_2008}, oblivating the need for additional finite size corrections. We also note that the chosen concentration almost agrees with the experimental proton concentration (1 mol L$^{-1}$ for $\mathrm{H^{+}}$).
As depicted in Fig.~\ref{fig1} (see the green arrow), the local potential gap $\left< \Delta \phi \right>_{\lambda}$ is then determined by:
\begin{align}
 e \left< \Delta \phi \right>_{\lambda} &= \mu - \left< \epsilon_\mathrm{1s,slab} \right> + \left< \epsilon_\mathrm{1s,bulk} \right>_{\lambda}. \label{eq_7}
\end{align}
Although the equation is simple, statistically accurate computations of the potential gap across the water-vacuum interface require expensive million-step MD simulations to yield thousands of uncorrelated water slab structures (see details in Section S4\dag). This problem was solved in the previous study~\cite{Jinnouchi_npjComputMater_2024} by employing machine-learned (ML) force fields (FFs)~\cite{Jinnouchi_PRB_2019} that allow for orders of magnitude faster MD simulations while retaining the accuracy of the FP method.

\begin{figure*}
\centering
\includegraphics[width=1.0\textwidth,angle=0]{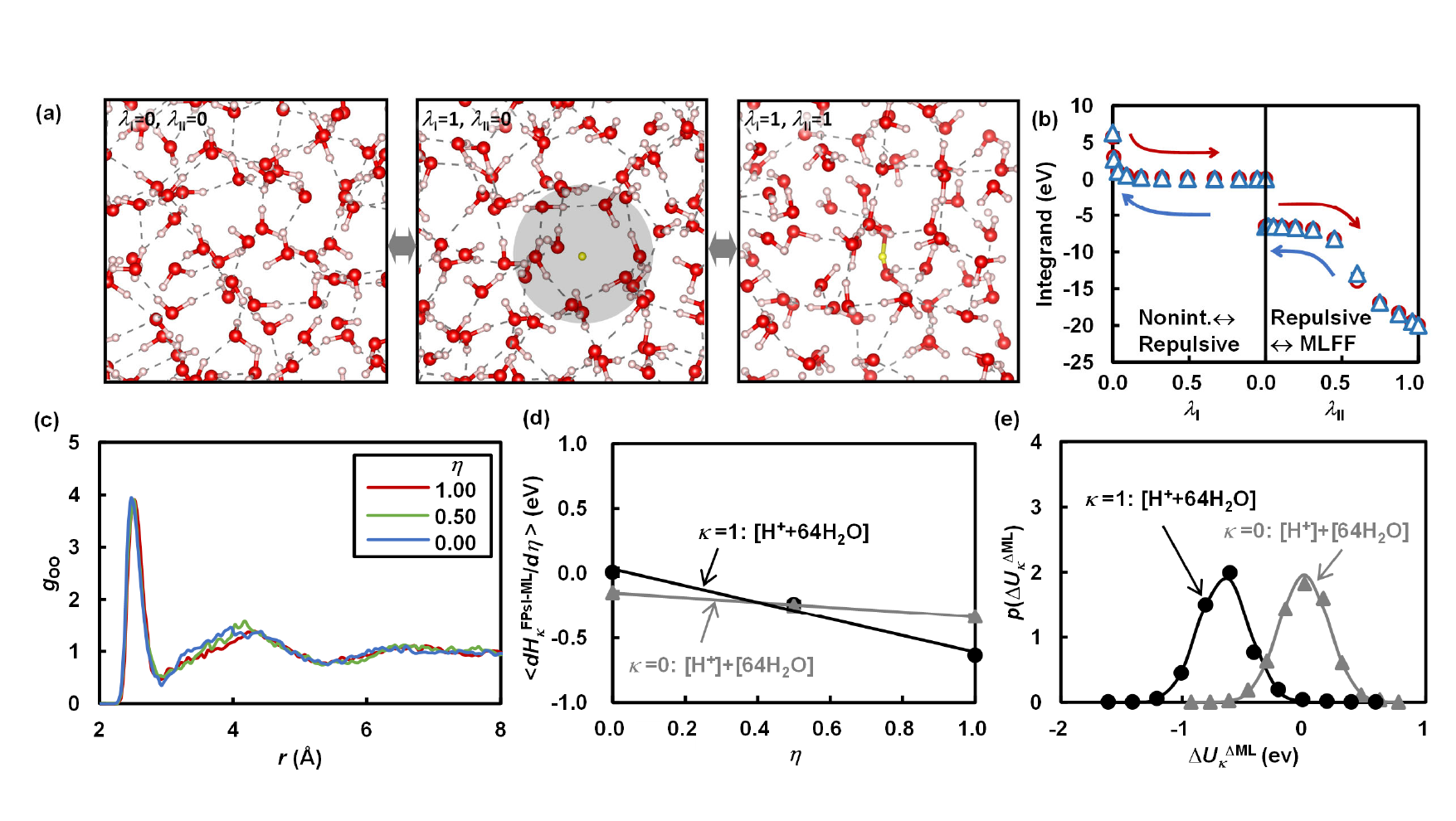}
\caption{Results of TI and TPT calculations: (a) Schematic of two TI steps from the non-interacting system ($\lambda_\mathrm{I}$=0 and $\lambda_\mathrm{II}$=0) to the system with a repulsive potential creating a cavity for proton placement ($\lambda_\mathrm{I}$=1 and $\lambda_\mathrm{II}$=0), and then to the fully interacting system described by MLFF ($\lambda_\mathrm{I}$=1 and $\lambda_\mathrm{II}$=1), (b) integrands of these TI steps along the coupling paths, (c) radial distribution functions between an oxygen atom in H$_{3}$O$^{+}$ and other oxygen atoms, (d) integrands in TI simulations along the coupling path $\eta$ from MLFF to RPBE+D3, and (e) probability distributions of the energy difference $\Delta U_{\mu}^\mathrm{\Delta ML}$ between PBE0+D3 and RPBE+D3 used in the TPT calculation. In (b), the red circles and blue triangles represent the integrands in the direction from the non-interacting proton to the interacting proton and vice versa, respectively. In (d) and (e), $\kappa=0$ corresponds to the gaseous proton  [H$^+$] and [water], 
and  $\kappa=1$ corresponds to the solvated system [H$^+$+ water].}
\label{fig2}
\end{figure*}

\subsection{Thermodynamic integration} 
The TI in the first term on the right-hand side of Eq.(\ref{eq_6}) is performed by the modified $\lambda$-MLFF scheme, which allows for insertion of atoms and molecules.~\cite{Jinnouchi_PRB_2020, Jinnouchi_JCP_2021} As in our previous study, the TI is decomposed into two steps: (1) TI using the MLFFs from the non-interacting proton and pure liquid water to the fully interacting aqueous solution containing the proton (or vice versa), and (2) TI from the MLFF potential to the FP potential. Step (1) is further decomposed into two steps (I) and (II) as shown in Fig.\ref{fig2} (a) to yield a reversible thermodynamic pathway between two endpoints. In the first step, a Gaussian soft-repulsive potential is gradually introduced between the proton and other atoms along the coupling parameter $\lambda_\mathrm{I}$ to form a cavity in water similar to the previous FP calculation of solvation free energies of Li$^{+}$ and F$^{-}$ ions.~\cite{Timothy_CS_2017} Subsequently, in the second step, the soft-repulsive potential is replaced with the full interaction represented by the MLFF model along the coupling parameter $\lambda_\mathrm{II}$. A point worth noting is that, unlike the TI approach used by Sprik and co-workers,~\cite{Cheng_JCP_2009, Costanzo_JCP_2011} no restraining potential is introduced in the final state where the MLFF is applied. 
For a system with 64 water molecules, we estimate that the neglected entropy contribution with a restraining potential is typically $1/\beta \ln (64) \approx 100 ~$ meV.
In our simulations, the free energy of the freely diffusing protons is accurately represented in the final state. It has been found that achieving a good convergence in the TI leading to fully diffusing protons requires simulations on the nanosecond scale. The MLFF make this possible by accelerating these MD simulations by several orders of magnitude. Details of these two TI simulations are explained in Section S2\dag. We refer to this improved method as the soft landing $\lambda$-MLFF scheme.  

While the TI step (1) provides a statistically accurate free energy for the MLFF, the MLFF model may introduce a non-negligible error. This error is corrected by the TI step (2). In this step, the potential energy transitions seamlessly from the MLFF to the FP potential. One of the main advantages of our ML-aided scheme is that the initial TI step from the non-interacting to the interacting system is done using the MLFF model. This initial step, which requires the use of an infinitesimally small interaction between the inserted atom and other atoms, becomes extremely challenging as the yet non-interacting atom approaches or even overlaps with other atoms and hence experiences a significant repulsive potential. Additionally, it is highly challenging to perform adequate configurational sampling of the proton, which diffuses through the Grotthuss mechanism, requiring multiple several ns to tens of ns of simulations along the coupling parameter with the FP method. The MLFF significantly accelerates these computations by several orders of magnitude. Another major advantage is that the accurate reproduction of FP structures by the MLFF results in small and nearly linear integrands with respect to the coupling parameter for the TI step (2), facilitating the convergence of this integral after a few tens of ps of MD simulations, thus reducing the high computational costs associated with FP calculations.

However, directly applying the final TI step (2) to an expensive FP method, like the hybrid functional used in this study, remains highly costly. Therefore, as in our previous study,~\cite{Jinnouchi_npjComputMater_2024} we correct the free energy calculated with the computationally inexpensive FP method, that is a semi-local functional in our case, through using thermodynamic perturbation theory (TPT) and a scheme akin to $\Delta$-ML~\cite{Balabin_JCP_2009, Ramakrishnan_JCTC_2015, Bartok_PRB_2013, Chmiela_NatCommun_2018, Sauceda_JCP_2019, Liu_2022_PRB, Carla_PRM_2023, Liu_PRL_2023}. Specifically the free energy
difference is calculated using:
\begin{align}
\Delta F &= F_1 - F_0 = -\frac{1}{\beta}\mathrm{ln} \left< e^{-\beta \Delta U} \right> _0  = -\frac{1}{\beta}\mathrm{ln} \left< e^{\beta \Delta U} \right> _1, \label{eq_8}
\end{align}
where the symbol $\Delta U$ denotes the potential energy difference between the expensive and inexpensive FP methods. 
Equation~(\ref{eq_8}) is, in principle, exact, but if the ensembles generated by the two potentials are too different, it may be necessary to evaluate the potential energy difference for thousands or even many ten thousands of configurations. 
Crucially, we suggest to calculate  the difference between the expensive  and inexpensive method $\Delta U$  using a ML representation.
Since the difference between any two first-principles methods is very smooth, only a few dozen structures are required to learn the difference 
($\Delta$-ML), as demonstrated in our previous study,~\cite{Jinnouchi_npjComputMater_2024}. Hence the ensemble averages in  Eq. (\ref{eq_8}) can
be calculated using thousands of  configurations at little cost.

\subsection{Correction to vibrational free energy}
Another obstacle in computational studies is the integration of nuclear quantum effects, which cannot be accounted for in the classical thermodynamic integration (TI) and transition path theory (TPT) simulations described previously. Specifically, the contributions of zero-point energy cannot be neglected due to the high frequency of the O-H bond.~\cite{Cheng_JCP_2009, Costanzo_JCP_2011, Ambrosio_JCP_2015, Ambrosio_JPCL_2018} In this study, nuclear quantum effects were estimated to be 0.30 eV by calculating the free energy difference between the quantum harmonic oscillator model and the classical model for three experimentally measured vibrational frequencies of 1250, 1760, and 3020 cm$^{-1}$,~\cite{Kim_JCP_2002} which were attributed to the solvation of a single proton by water (details in Section S3\dag). Our estimated value is slightly smaller than the previously reported value of 0.36 eV,~\cite{Ambrosio_JCP_2015} which was calculated solely from the contribution of zero-point energies, excluding entropy contributions and the subtraction of classical vibrational contributions. The discrepancy, as explained in Section S3\dag, stems from whether these additional contributions are accounted for. To examine the sensitivity of the correction to the model and frequencies, the correction was also calculated for H$_{3}$O$^{+}$ and H$_{2}$O molecules in vacuum. The results indicate that the correction is insensitive to details of the model and frequencies (see also Section S3\dag).

\subsection{Redox potentials of other electron transfer reactions} 
Finally, to demonstrate that the hybrid functional consistently predicts the redox potentials of a wide range of redox couples with high precision, we calculated the redox potentials for electron transfer reactions, Ox + e$^{-}\leftrightarrow$Red, of three redox couples, V$^{3+}$/V$^{2+}$, Ru$^{3+}$/Ru$^{2+}$, and O$_{2}$/O$_{2}^{-}$, in addition to the three redox couples, Fe$^{3+}$/Fe$^{2+}$, Cu$^{2+}$/Cu$^{+}$, and Ag$^{2+}$/Ag$^{+}$, that were calculated in our previous study.~\cite{Jinnouchi_npjComputMater_2024} The redox potentials were calculated from the free energy changes computed via the ML-aided TI and TPT simulations from the oxidized state to the reduced state. Details of the computational method are explained in our previous publication.~\cite{Jinnouchi_npjComputMater_2024} Details of the models are described in Section S1\dag.

\subsection{Computational method} 
All simulations were carried out using the Vienna Ab initio Simulation Package (VASP).~\cite{Kresse_PRB_1996,Kresse_CMS_1996, Kresse_PRB_1999} For the MLFF models, we utilized the algorithm described in our previous publications.~\cite{Jinnouchi_PRB_2019, Jinnouchi_JPCL_2023} Similar to the pioneering ML approaches,~\cite{Behler_PRL_2007, Bartok_PRL_2010} the potential energy is approximated as a sum of local energies, which are further approximated used a spare representation of the kernels as a weighted sum of kernel basis functions. The Bayesian framework allows for accurate predictions of energies, forces, and their uncertainties, enabling efficient on-the-fly sampling of reference structures during FPMD simulations of target systems. For the proton insertion calculations performed to compute ASHEP, a single MLFF was trained on both the reactant and product states. Additionally, to generate a stable and reversible thermodynamic pathway, the MLFF was also trained on the fly during the TI simulation along the coupling constant $\lambda_\mathrm{II}$ prior to the production run. For other electron transfer reactions, individual MLFF models were trained separately on the reactant and product states and were further trained on the fly during the TI simulation along the coupling constant.
Exchange-correlation interactions between electrons were modeled using the modified Perdew-Burke-Ernzerhof semi-local functional (RPBE+D3),~\cite{Hammer_PRB_1999} augmented with Grimme's dispersion interaction.~\cite{Grimme_JCP_2010, Grimme_JCC_2011} Details of the equations, parameters, and training conditions are summarized in Section S1\dag. The same formulation was applied to the $\Delta$-ML model. A $\Delta$-ML model was generated for each reactant and product state. The training was conducted on the differences in energies and forces between the semi-local RPBE+D3 functional and the non-local hybrid functional, which includes 25\% exact exchange with and without Grimme's dispersion interaction (PBE0 and PBE0+D3).~\cite{Adamo_JCP_1999} These functionals were shown to reproduce the redox potentials of three transition metals: Fe, Cu, and Ag, in our previous study.~\cite{Jinnouchi_npjComputMater_2024}

\begin{figure}[t]
\includegraphics[width=0.44\textwidth,angle=0]{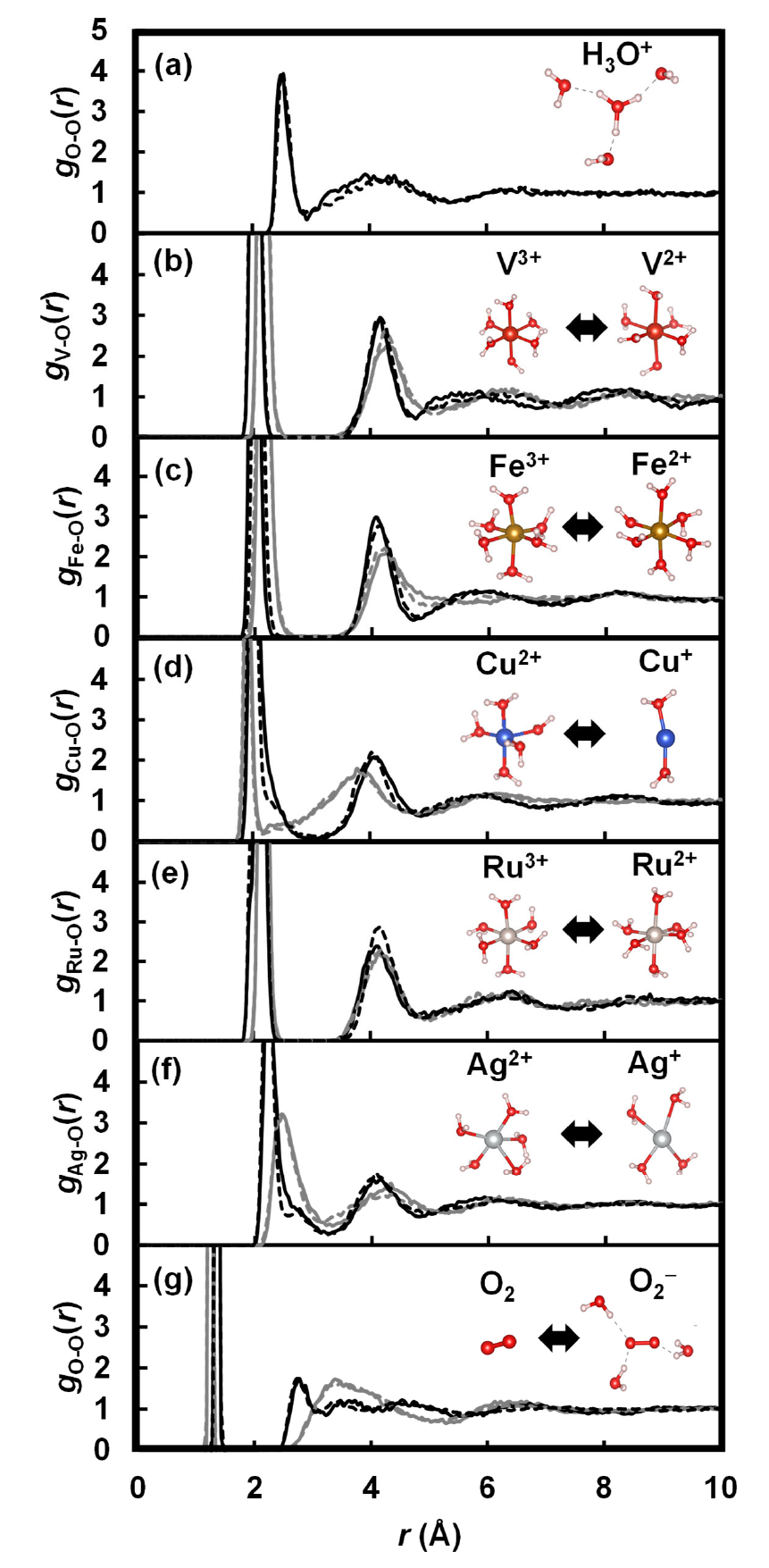}
\caption{Radial distribution functions between (a) O in H$_{3}$O$^{+}$ and other oxygen atoms in the H$^{+}$+64H$_{2}$O system,
 (b) V and O in the V$^{3+}$/V$^{2+}$+64H$_{2}$O systems, (c) Fe and O in the Fe$^{3+}$/Fe$^{2+}$+64H$_{2}$O systems, (d) Cu and O in the Cu$^{2+}$/Cu$^{+}$+64H$_{2}$O systems, (e) Ru and O in the Ru$^{3+}$/Ru$^{2+}$+64H$_{2}$O systems, (f) Ag and O in the Ag$^{2+}$/Ag$^{+}$+64H$_{2}$O systems, and (g) O in O$_{2}$/O$^{-}_{2}$ and other oxygn atoms in the O$_{2}$/O$^{-}_{2}$+64H$_{2}$O systems. Solid and dashed lines are the RDFs obtained by the MLFF models and the FP method, respectively. Black and gray lines in (a) to (g) show the RDFs of the oxidized state and reduced state, respectively.}
\label{fig3}
\end{figure}

\section{\label{section3}Results and discussion}

\subsection{Accuracy of MLFF and $\Delta$-ML models}
The MLFF models for seven redox couples achieve root mean square errors (RMSEs) of 1-2 meV/atom, 40-80 meV/Å, and 0.40-1.1 kbar, as shown in Table S2\dag (error distributions are also presented in Figures S2 to S8\dag). These RMSEs are comparable to those of MLFFs in previous studies.~\cite{Jinnouchi_PRL_2019, Jinnouchi_PRB_2019, Jinnouchi_PRB_2020, Jinnouchi_JCP_2020} Due to their accuracy, the MLFFs  reproduce the solvation structure surrounding the redox couples, as demonstrated by the radial distribution functions (RDFs) shown in Figure~\ref{fig3}. Consistent with the previous FPMD simulation,~\cite{Tuckerman_JCP_1995} H$^{+}$ forms an H$_{3}$O$^{+}$ ion coordinated by three water molecules on average. Similarly, Fe and Ru form rigid first solvation shells composed of six water molecules, regardless of the oxidation state, in alignment with previous computational results.~\cite{Remsungen_CPL_2004, Blumberger_JPCB_2005, Jaque_JPCC_2007, Bogatko_JPCA_2010} Vanadium also forms a rigid first solvation shell composed of six water molecules. In contrast, as reported in previous studies,~\cite{Blumberger_JACS_2004, Jinnouchi_npjComputMater_2024} the coordination number for Cu changes from 5-6 in the oxidized state (Cu$^{2+}$) to 1-2 in the reduced state (Cu$^{+}$), and for Ag, it changes from 5-6 in the oxidized state (Ag$^{2+}$) to 4-5 in the reduced state (Ag$^{+}$). Notable H-bonds are not formed between the neutral O$_{2}$ and other water molecules, while its reduced state, O$^{-}_{2}$, is coordinated by roughly three water molecules on average. Although the MLFF models reproduce the solvation structures observed in our FPMD calculations and previous calculations, non-negligible deviations are present in the RDFs calculated by the MLFF models compared to those calculated by the FP method. The error in the MLFF models can be corrected through the TI simulation from the MLFF potential to the FP potential, as illustrated by the transition of the radial distribution function (RDF) shown in Fig.~\ref{fig2} (c). 

Thanks to the smooth energy difference between the semi-local and hybrid functionals, the $\Delta$-ML models accurately represent the energy difference with only 40 configurations, achieving RMSEs more than an order of magnitude smaller than those of the MLFFs, as shown in Table S3 and Figures S2 to S8\dag. 
Due to this remarkable accuracy,  the $\Delta$-ML models can provide exceedingly accurate free energy changes via TPT simulations without further correction.\cite{Jinnouchi_npjComputMater_2024}

\begin{figure}
\centering
\includegraphics[width=0.40\textwidth,angle=0]{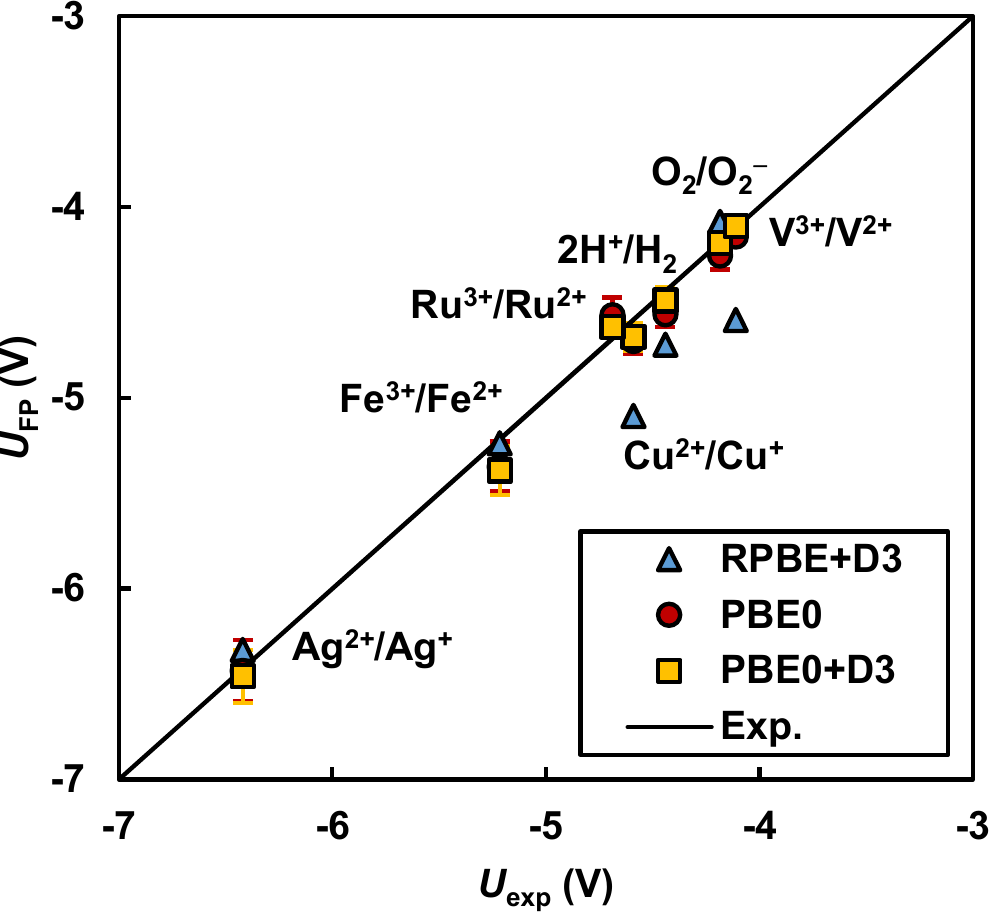}
\caption{Comparison of the absolute redox potentials of seven redox couples determined by three functionals with their experimental values. Data for transition metals, Fe$^{3+}$/Fe$^{2+}$, Cu$^{2+}$/Cu$^{+}$, and Ag$^{2+}$/Ag$^{+}$, were taken from our previous publication.~\cite{Jinnouchi_npjComputMater_2024}}
\label{fig4}
\end{figure}

\subsection{Real potential of proton}
As illustrated in Fig.\ref{fig2}(b), thermodynamic integration (TI) simulations using the MLFF trained on the semi-local RPBE+D3 functional generate smooth and reversible integrands along the coupling parameters, within the range of statistical errors. At the endpoint of TI ($\lambda_\mathrm{I}=1$ and $\lambda_\mathrm{II}=1$), as shown in Fig.~\ref{fig2}(c), the MLFF well reproduces the RDF calculated using the RPBE+D3 functional. Owing to this high precision, as shown in Fig.~\ref{fig2}(d), TI from MLFF to the RPBE+D3 functional yields nearly linear and relatively small integrands along the coupling path $\eta$. Consequently, this TI induces only slight modifications to the real potential of the proton ($\alpha_\mathrm{H^{+}}^{0}$) (40 meV). 
The calculated value shown in Table~\ref{table1} ($-$11.02$\pm$0.06 eV) is already close to the experimental result ($-$11.28$\pm$0.02 eV) and also closely matches a previously calculated value of $-$11.00 eV. This earlier calculated value was derived by adding 0.08 eV to the tabulated value of W$_\mathrm{H^{+}}=-11.08$ eV reported by Ambrosio et al.~\cite{Ambrosio_JPCL_2018} This adjustment was made to account for the dilution of the hydrogen atom in the gas phase from a concentration of 1 mol L$^{-1}$ to 0.1 MPa (1/24.46 mol L$^{-1}$), to make it comparable with Trasatti's value. The agreement with experiment is further improved through hybrid functional calculations facilitated by TPT simulations utilizing the $\Delta$-ML scheme. The TPT simulations demonstrate that the probability distributions of energy differences between the hybrid functional (PBE0 or PBE0+D3) and RPBE+D3 are accurately represented by Gaussian distributions, as depicted in Fig.~\ref{fig2}(e). This implies that the second cumulant expansion equation [Eq. (S9)\dag] provides a reasonable approximation. The computed value for PBE0+D3 ($-$11.12$\pm$0.09 eV) shows the best agreement with experiment.

\subsection{ASHEP and other redox potentials}
The ASHEP values calculated by RPBE+D3, PBE0, and PBE0+D3 are $-$4.75$\pm$0.05, $-$4.58$\pm$0.09, and $-$4.52$\pm$0.09 V, respectively. While error cancellation between the hydrogen dissociation free energy $\Delta_\mathrm{at} G^{0}$ and the other two properties is a partial cause, the PBE0+D3 functional shows excellent agreement with the IUPAC recommended experimental ASHEP value of $-$4.44$\pm$0.02 V. Furthermore, the PBE0+D3 functional accurately reproduces experimental values of the redox potentials of five transition metal redox couples, Ag$^{2+}$/Ag$^{+}$, Fe$^{3+}$/Fe$^{2+}$, Cu$^{2+}$/Cu$^{+}$, V$^{3+}$/V$^{2+}$, Ru$^{3+}$/Ru$^{2+}$, and the molecular redox couple O$_{2}$/O$^{-}_{2}$ as shown in Fig.~\ref{fig4} (all relevant data are in Fig. S9 and Table S5\dag). Comparison with experimental data demonstrates that the PBE0+D3 functional consistently reproduces experimental redox potentials across a broad range of potentials with a remarkably small RMSE of 80 mV.

\section{Conclusions}
We have developed an ML-aided FP method that allows for statistically accurate computation of the free energy change associated with proton insertion into aqueous solutions through TI and TPT calculations. In this approach, most of the energetic contributions to the free energy change were obtained through TI simulations using MLFF models trained on semi-local exchange-correlation functionals. Small residual errors in the MLFF models were corrected via subsequent TI simulations transitioning from the MLFF potential to the FP potential. Moreover, the $\Delta$-ML model, which learned the potential energy difference between semi-local and hybrid functionals, enabled efficient TPT calculations of free energy changes for the expensive hybrid functional. Overall, our scheme accelerates free energy computations by about two orders of magnitude, enabling the calculation of redox potentials for seven redox couples on an absolute scale, including the ASHEP. This application has shown that when combined with a hybrid functional (PBE0), the method can predict redox potentials with an exceptional average accuracy of 80 mV. The ML surrogate models that facilitate proton and electron insertions offer a highly flexible, accurate  and efficient framework for predicting free energy changes and redox potentials on an absolute scale, utilizing high-level electronic structure theories for a variety of proton and electron transfer reactions.

\section*{Data availability}
We have provided all relevant data in the ESI of the paper.


\section*{Conflicts of interest}
There are no conflicts of interest to declare.

\section*{Acknowledgement}
This research was funded in part by the Austrian Science Fund (FWF) 10.55776/COE5. For open access purposes, the author has applied a CC BY public copyright license to any author accepted manuscript version arising from this submission.

\section*{Code availability}

The VASP code is distributed by the VASP Software GmbH. The machine learning modules will be included in the release of vasp.6.3. Prerelease versions are available from G.K. upon reasonable request.

\nocite{*}
\bibliographystyle{naturemag}
\bibliography{./ms}

\end{document}